\documentclass[preprint,10pt]{aastex}

\slugcomment{{\it Publications of the Astronomical Society of the Pacific}, 113: 1349-1364, 2001 November}

\newcommand{\etal}{et al.\ }
\newcommand{\phmm}{\phm{$-$}}

\begin{document}

\title{$BVRI$ Photometry of Supernovae}

\author{Wynn~C.~G.~Ho\altaffilmark{1,}\altaffilmark{2},
Schuyler~D.~Van~Dyk\altaffilmark{3},
Chien~Y.~Peng\altaffilmark{4}, Alexei~V.~Filippenko\altaffilmark{5},
Douglas~C.~Leonard\altaffilmark{6}, Thomas~Matheson\altaffilmark{7}, 
Richard~R.~Treffers\altaffilmark{8}}

\affil{Department of Astronomy, 601 Campbell Hall, University of California,
Berkeley, CA 94720-3411}
\and
\author{Michael~W.~Richmond\altaffilmark{9}}
\affil{Department of Physics, Rochester Institute of Technology,
Rochester, NY 14623-5603}

\altaffiltext{1}{Also associated with Space Sciences Laboratory, and Department
of Physics, University of California, Berkeley.}
\altaffiltext{2}{Current address: Department of Astronomy,
Cornell University, Ithaca, NY 14853; e-mail wynnho@astro.cornell.edu.}
\altaffiltext{3}{Current address: IPAC, 100-22 Caltech, Pasadena, CA 91125;
e-mail vandyk@ipac.caltech.edu.}
\altaffiltext{4}{Current address: Steward Observatory, University of Arizona,
Tucson, AZ 85721; e-mail cyp@as.arizona.edu.}
\altaffiltext{5}{E-mail alex@astro.berkeley.edu.}
\altaffiltext{6}{Current address: Department of Astronomy, University of
Massachusetts, Amherst, MA 01003-9305; e-mail leonard@nova.astro.umass.edu.}
\altaffiltext{7}{Current address: Harvard/Smithsonian Center for Astrophysics,
60 Garden St., Cambridge, MA 02138; e-mail tmatheso@cfa.harvard.edu.}
\altaffiltext{8}{E-mail treffers@pacbell.net.}
\altaffiltext{9}{E-mail mwrsps@rit.edu.}

\begin{abstract}
We present optical photometry of one Type~IIn supernova (1994Y) and nine
Type~Ia supernovae (1993Y, 1993Z, 1993ae, 1994B, 1994C, 1994M, 1994Q, 1994ae,
and 1995D).  SN~1993Y and SN~1993Z appear to be normal SN~Ia events with
similar rates of decline, but we do not have data near maximum brightness.  The
colors of SN~1994C suggest that it suffers from significant reddening or is
intrinsically red.  The light curves of SN~1994Y are complicated; they show a
slow rise and gradual decline near maximum brightness in $VRI$ and numerous
changes in the decline rates at later times.  SN~1994Y also demonstrates color
evolution similar to that of the SN~IIn~1988Z, but it is slightly more luminous
and declines more rapidly than SN 1988Z.  The behavior of SN~1994Y indicates a
small ejecta mass and a gradual strengthening of the H$\alpha$ emission
relative to the continuum.
\end{abstract}

\keywords{supernovae: general -- supernovae: individual (1993Y, 1993Z,
1993ae, 1994B, 1994C, 1994M, 1994Q, 1994Y, 1994ae, 1995D)}

\section{Introduction}

  Supernova (SN) light curves reveal the structure of the stellar progenitors
and reflect the underlying energy sources created in the explosion. They have
also been the primary tool for intercomparison of individual events and
selection of subsamples of supernovae (SNe) for use as distance indicators.
Direct comparisons of peak luminosities and multiband light curves with the
predictions of theoretical models lead to a better understanding of supernova
explosions.  Useful reviews of the optical light curves of SNe are given by
Doggett \& Branch (1985), Kirshner (1990), Wheeler \& Harkness (1990), Patat
\etal (1994), Leibundgut (1996), and Suntzeff (1996), among others.

 We have been engaged in systematic optical monitoring of bright, relatively
nearby SNe. These include hydrogen-rich objects (Type~II), as well as the
various kinds of hydrogen-deficient SNe (Types~Ia, Ib, and Ic).  (For a summary
of the spectroscopic properties of SNe, see Filippenko 1997.) Here we report
our results for nine SNe~Ia and one peculiar SN~II, from data obtained in
1993--1995.  Section 2 discusses the discoveries of the SNe, and Section 3
contains a description of the observations.  The ``template subtraction''
method of photometry used to extract the light of SNe from their environs is
presented in Section 4.  Section 5 describes the calibration of the
measurements onto the standard Johnson-Kron-Cousins system.  The SN light
curves and
color curves appear in Sections 6 and 7, respectively, while estimates of the
absolute magnitudes are in Section 8.  Section 9 summarizes our results.  UT
dates are used in all cases.

\section{Discoveries}

The discoveries of the supernovae presented here were made by several groups.
SN~1993Y (Fig. 1) in UGC~2771 was discovered on 1993 September 18, about $9''$
east and $39{\farcs}6$ north of the galaxy's nucleus, with the 1.2-m
Oschin telescope as part of the second Palomar Sky Survey (Mueller 1993).
Accurate position measurements of the object resulted in \mbox{$\alpha_{1950} =
3^{h}28^{m}6^{s}.75$}, \mbox{$\delta_{1950} = +39^{\circ}34'51{\farcs}9$}
(Balam 1993).  A spectrum taken on September 22 showed SN~1993Y to be of
Type~Ia about three weeks past maximum brightness (Dressler \& Sargent 1993).
Another spectrum taken on September 25 showed that it was close to one month
past maximum (Filippenko \& Matheson 1993).

SN~1993Z (Fig. 2) in the Sab galaxy NGC~2775 was found on 1993 September 23,
about $15''$ west and $42''$ south of the nucleus, as part of the Leuschner
Observatory Supernova Search (LOSS) using a 0.76-m telescope (Treffers
\etal 1993).  The accurate position measurement of the SN is
\mbox{$\alpha_{1950} = 9^{h}7^{m}40^{s}.18$}, \mbox{$\delta_{1950} =
+7^{\circ}13'56{\farcs}1$} (Balam 1993).  A spectrum taken on September 25
indicated SN~1993Z to be of Type~Ia about four weeks past maximum (Treffers
\etal 1993).

SN~1993ae (Fig. 3) in UGC~1071 was found with the 0.9-m Schmidt
telescope at the Observatoire de la Cote d'Azur (OCA) on 1993 November 7.92;
the SN was located at \mbox{$\alpha_{1950} = 1^{h}27^{m}16^{s}.08$},
\mbox{$\delta_{1950} = -2^{\circ}14'5{\farcs}5$} (Pollas 1993).  Spectroscopy
on November 11.2 showed that SN~1993ae was a normal Type~Ia about 10~d after
maximum with an expansion velocity of about 10000~km~s$^{-1}$ from
Si~II~$\lambda$6355 (Cappellaro \& Bragaglia 1993).  The supernova was located
about $16''$ east and $24''$ north of the nucleus.

SN~1994B (Fig. 4) was discovered on 1994 January 16.98 in an anonymous galaxy,
again using the OCA 0.9-m Schmidt telescope; the supernova position was
at \mbox{$\alpha_{1950} = 8^{h}17^{m}51^{s}.64$}, \mbox{$\delta_{1950} =
15^{\circ}53'20{\farcs}5$}, or $4{\farcs}2$ east and $3{\farcs}9$ north of the
center of the host face-on spiral galaxy (Pollas 1994).  A spectrum obtained on
January 19 suggested that SN~1994B was of Type~Ia near maximum; the broad
Si~II~$\lambda$6355 absorption trough was visible, and a redshift $z = 0.089$
of the parent galaxy was also determined from the H~II region on which SN~1994B
was superposed (Filippenko \& Matheson 1994).

SN~1994C (Fig. 5) was discovered on 1994 March 5 at \mbox{$\alpha_{1950} =
7^{h}53^{m}8^{s}.7$}, \mbox{$\delta_{1950} = 45^{\circ}0'20{\farcs}8$} during
the second Palomar Sky Survey (Mueller \& Mendenhall 1994).  It was confirmed
as a SN on March~9 (Djorgovski, Thompson, \& Smith 1994).  SN~1994C was
classified on March~12 as a Type~Ia about 20~d past maximum, and the host
galaxy's redshift was determined to be $z = 0.051$ (Riess, Challis, \& Kirshner
1994).  It was located about $3''$ west and $6''$ north of the anonymous
galaxy's nucleus.

SN~1994M (Fig. 6) in NGC~4493 was found by Wild (1994) at Zimmerwald,
University of Berne, on 1994 April 29.9; the SN position was
\mbox{$\alpha_{1950} = 12^{h}28^{m}35^{s}.1$}, \mbox{$\delta_{1950} =
0^{\circ}52'54''$}, about $3''$ west and $28''$ south of the nucleus.  A
spectrum taken on May 4.14 showed it to be of Type~Ia very close to maximum
light (Schmidt, Kirshner, \& Peters 1994).

SN~1994Q (Fig. 7) was found with the OCA 0.9-m Schmidt telescope on 1994
June 2.92 at \mbox{$\alpha_{1950} = 16^{h}48^{m}11^{s}.64$},
\mbox{$\delta_{1950} = +40^{\circ}31'1{\farcs}1$}, or $0{\farcs}4$ west and
$3{\farcs}5$ south of an anonymous galaxy's nucleus (Pollas \& Albanese 1994).
A spectrum taken on June~3 suggested that SN~1994Q was of Type~Ia about one
week past maximum (Filippenko, Matheson, \& Barth 1994).

SN~1994ae (Fig. 8) in the Sc galaxy NGC~3370 was discovered on 1994 November 14
by LOSS (Van Dyk \etal 1994) long before maximum light; various estimates of
the date of maximum visual brightness were November 30 (Nakano \etal 1994),
December 1 (Patat, Vician, \& Szentasko 1994), and prior to December 6
(Vanmunster, Villi, \& Cortini 1994). The accurate position is
\mbox{$\alpha_{1950} = 10^{h}44^{m}21^{s}.52$}, \mbox{$\delta_{1950} =
+17^{\circ}32'20{\farcs}7$} (Nakano, Kushida, \& Kushida 1994), which is
$30{\farcs}3$ west and $6{\farcs}1$ north of the galaxy nucleus. A spectrum
taken on November 23.16 indicated SN~1994ae to be of Type~Ia; the measured
expansion velocity for the Si~II~$\lambda$6355 line was 11000~km~s$^{-1}$
(Iijima, Cappellaro, \& Turatto 1994).

SN~1995D (Fig\ 9) in the S0 galaxy NGC~2962 was found apparently on the rise on
1995 February 10.76, using a 0.25-m telescope at the Yatsugatake South
Base Observatory (Kushida, Nakano, \& Kushida 1995) at \mbox{$\alpha_{2000} =
9^{h}40^{m}54^{s}.79$}, \mbox{$\delta_{2000} = +5^{\circ}8'26{\farcs}6$}, about
$11''$ east and $90{\farcs}5$ south of the galactic nucleus.  A spectrum
obtained on February 13.18 confirmed it as a Type~Ia, about one week before
maximum; the spectrum was dominated by P-Cygni lines of intermediate-mass
elements on a very blue continuum, and the deduced expansion velocity for
Si~II~$\lambda$6355 was 10900~km~s$^{-1}$ (Benetti, Mendes de Oliveira, \&
Manchado 1995).

SN~1994Y (Fig. 10) in the Sbc galaxy NGC~5371 was discovered on 1994 August
19.15 with a 0.9-m telescope at McDonald Observatory (Wren 1994).
Images obtained as part of the LOSS showed it to be about $34''$ west and
$11''$ north of the nucleus, on the rise from August 12 to August 16, and not
visible on August 3, 6, and 9 (Paik \etal 1994).  A spectrum taken on August 25
identified it as a Type~II (Jiang, Liu, \& Hu 1994).  Furthermore, Clocchiatti \etal
(1994) took a spectrum on August 27.15, which showed the Balmer emission lines
without the characteristic P-Cygni profiles; rather, there were narrow lines on
top of broad bases, thereby resulting in the classification of SN~1994Y as a
Type~IIn (see Filippenko 1997).  Also evident in a spectrum taken on 1995
February 8 was a narrow absorption feature at 6622~\AA\ superimposed on the
strong H$\alpha$ line, signifying a large amount of circumstellar material
(Wang \etal 1995).  Accurate position measurements yielded \mbox{$\alpha_{1950}
= 13^{h}53^{m}30^{s}.57$}, \mbox{$\delta_{1950} = +40^{\circ}42'32{\farcs}8$}
(Boattini \& Tombelli 1994).

\section{Observations}

Observations were made at Lick Observatory with the Nickel 1-m reflector and
standard Johnson-Kron-Cousins $BVRI$ filters.  The telescope utilizes a Loral
$2048 \times 2048$ pixel CCD (15 $\mu$m pixel$^{-1}$), which we binned
$2 \times 2$ to yield a plate scale of $0{\farcs}37$ pixel$^{-1}$.  The
full-width at half-maximum (FWHM) of stellar objects generally varied from
about $1{\farcs}5$ to $2''$.  The CCD images were bias corrected using an
average of several bias images.  They were also flattened by the average of
several flatfield images of the twilight sky.  Exposure times were as short as
120~s for SN~1994ae near maximum light to 1800~s for deeper late-time and
template images.  However, some images were combined to give a longer overall
exposure time.  Figures 1--10 illustrate Lick images of the ten SNe, their host
galaxies, and the stars in the field that were used as comparisons (henceforth
referred to as field stars).

SN~1993Y, SN~1993Z, and SN~1994Y were also observed at Leuschner Observatory as
part of the Berkeley Automatic Imaging Telescope (BAIT) project, which employed
50-cm and 76-cm telescopes (see Filippenko 1992; Richmond, Treffers, \&
Filippenko 1993; Richmond \etal 1996; and references therein). Each telescope
was equipped with a $512 \times 512$ pixel CCD with a plate scale of about
$0{\farcs}65$ pixel$^{-1}$, and stellar objects generally had a FWHM of about
$3''$ to $4''$.  The filters were made according to the characteristics
described by Bessell (1990).  The images were bias corrected with the median of
five bias images taken each afternoon.  Five flat-field images of the twilight
sky were used to determine a median image, which was then used to flatten the
data.  Exposure times were between 120 and 900~s.  SN~1993Y was observed from
1993 September 26 through December 20, SN~1993Z from 1993 September 23 through
1994 February 1, and SN~1994Y from 1994 August 22 through December 7.

In addition, three images of SN~1994Y were taken at Kitt Peak National
Observatory on 1995 January 28, using a 0.9-m telescope and Tektronix $2048
\times 2048$ pixel CCD (24 $\mu$m pixel$^{-1}$).  They were each 300-s
exposures at a plate scale of $0{\farcs}68$ pixel$^{-1}$ and FWHM from $2''$
to $3''$.

\section{Photometry}

To determine the instrumental magnitudes of most of these various SNe, we use
the technique of template subtraction described in Richmond \etal (1995; see
also Filippenko \etal 1986) to minimize the light contamination by the
background galaxy.  This involves taking a ``template'' image of the galaxy
long after the supernova has faded.  The template image was matched to each
early-time image by registering the template using several field stars in
common.  The template was obtained under superior seeing conditions and was
then convolved with a Gaussian to match the widths of the point-spread function
(PSF) of each early-time image.  The magnitudes of the field stars for both the
template and early-time images were measured using a circular aperture of 15
pixels radius and a sky annulus of width 10 pixels.  Comparing the magnitudes,
an average scale factor was obtained and used to subtract the template image
from the early-time image.  The flux scale factor varied by $\le$ 10\%
typically from star to star within a given frame.  The resulting
template-subtracted image showed only the SN and, occasionally, the small
residuals of the field stars and the host galaxy nucleus.  The SN magnitude was
then measured with negligible contamination from the galaxy light.

For SNe~1993ae, 1994M, and 1995D, template subtraction was not needed, because
the SN was sufficiently far from the galactic nucleus and in a region of
relatively uniform background.  Simple aperture photometry was performed in
these cases, with a circular aperture of 10 or 15 pixels radius and a
10-pixel-width sky annulus.

PSF-fitting photometry, via IRAF\footnote{IRAF (Image Reduction and Analysis
Facility) is distributed by the National Optical Astronomy Observatories, which
are operated by the Association of Universities for Research in Astronomy,
Inc., under cooperative agreement with the National Science
Foundation.}/DAOPHOT (Stetson 1987), was performed for SN~1994Y and for the
Leuschner observations of SN~1993Y.  In both cases, the model PSF was
constructed using bright, isolated field stars on each of the images.  SN~1993Y
was far enough from the host galaxy nucleus that the background was negligible
in all bands.

\section{Calibrations}

On particular nights when the conditions were photometric, standard star fields
from Landolt (1992) were observed in $BVRI$, along with images of the SN.  The
instrumental magnitudes of these stars were determined using either a 10- or
15-pixel-radius circular aperture and a 10-pixel-width sky annulus.  Using
IRAF, the photometric instrumental magnitudes were converted to apparent
magnitudes with the transformation equations
\begin{eqnarray}
b & = & (B - V) + V + \mbox{b}_1 + \mbox{b}_2 X_b + \mbox{b}_3 (B - V)
 \label{eq:b} \\
v & = & V + \mbox{v}_1 + \mbox{v}_2 X_v + \mbox{v}_3 (B - V)
 \label{eq:v} \\
r & = & V - (V - R) + \mbox{r}_1 + \mbox{r}_2 X_r + \mbox{r}_3 (V - R)
 \label{eq:r} \\
i & = & V - (V - R) - (R - I) + \mbox{i}_1 + \mbox{i}_2 X_i
 + \mbox{i}_3 (R - I), \label{eq:i}
\end{eqnarray}
where $bvri$ are instrumental magnitudes, $BVRI$ are standard apparent
magnitudes, and $X$ are the airmass values.  The second-order extinction terms
were set to zero.  The results, along with the number of Landolt fields
observed and used on each night to determine the photometric solution, are
shown in Table~1.  In solving these equations, the standard deviations in the
residuals were on the order of a few hundredths of a magnitude.  Thus, the net
uncertainty of the fit can be taken to be $\leq 0.03$ mag.

In each of Figures 1--10, the labeled field stars were the comparisons used to
determine the apparent magnitude of the respective SN.  Each field star's
instrumental magnitude was measured, again using the same aperture and sky
annulus as above, and transformed with Equations (1)--(4) into apparent
magnitudes.  Table 2 lists the results and their respective uncertainties.  The
images were fairly clean, so that cosmic ray hits near the stars of interest
occurred only rarely.  In such cases where the cosmic-ray pixels could be
easily identified, they were smoothed by linear interpolation using nearby
pixels.  In cases when there were no photometric observations made, the
apparent magnitudes of the field stars were obtained from Riess (1996).  Also,
observations of SNe~1993Z, 1994ae, 1995D, and 1994Y occurred on multiple nights
under photometric conditions; the night with the smallest photometric
uncertainties was adopted as the standard.  Comparisons of our field star
calibrations with Riess (1996) and Riess \etal (1999) were made, whenever
possible, and the results were generally quite consistent.

Each SN instrumental magnitude was then converted into an apparent magnitude on
the standard Johnson-Kron-Cousins $BVRI$ system using a weighted mean of the
difference between the instrumental and apparent magnitudes of the field stars.
Tables~3--6 present these magnitudes along with their uncertainties.  In
addition, a comparison of the SN~1994Y data using PSF fitting versus
template-subtraction yielded negligible differences when the SN was bright.
Comparisons with the results of Riess (1996) and Riess \etal (1999) generally
show good agreement for the objects and epochs in common: the mean difference
for 34 points is $\sim$ 0.01 mag, and the dispersion is $\sim$ 0.07 mag.
However, there may be a slight systematic trend, namely our average values are
$\sim$ 0.04, 0.02, 0.01 mag brighter in $B$, $V$, and $R$, respectively, and
$\sim$ 0.03 mag fainter in $I$ than those of Riess (1996) and Riess et
al. (1999).

Finally, we note that the transformation equations (\ref{eq:b})-(\ref{eq:i}),
which were used on photometric nights at Lick to derive apparent magnitudes of
the field stars from their instrumental magnitudes, are implicitly assumed to
be valid for the Leuschner and KPNO data.  From the apparent magnitudes of the
field stars, we find the scaling relation for the nights that were not
photometric and scale the SN instrumental magnitude to an apparent magnitude.
Although the Loral chips were changed in the middle of the Lick program, the
two CCDs were very similar (both Loral) and the filters remained the same, so
nearly the same transformation should apply.  The lack of independent
transformations for the Leuschner and KPNO data increases the uncertainties of
these data but by an amount that is difficult to quantify. Unfortunately, the
Leuschner system that was used is no longer available for calibration; however,
we still obtain consistent results between Lick and Leuschner. There were only
three KPNO measurements, and again, they are consistent with the other results.

\section{Optical Light Curves}

\subsection{Type Ia Supernovae}

Figure 11 shows the light curves of SN~1993Y, and Figure 12 shows the light
curves of SN~1993Z.  The light curves of SNe~1993ae, 1994B, 1994C, 1994M,
1994Q, 1994ae, and 1995D are shown in Figure 13.  The line segments connecting
the points have no significance other than to aid the reader in distinguishing
between different SNe and to indicate general trends.

The time of maximum light for SN~1993Y is uncertain.  Based on its spectra, it
was estimated to be either around August 26 (Filippenko \& Matheson 1993) or
September 1 (Dressler \& Sargent 1993).  For this paper, we adopt August 26 as
the date of peak brightness.  The observations of SN~1993Y are therefore from
31~d to 137~d past maximum.  Normal Type~Ia SNe show a decrease in the decline
rate to a slow linear magnitude decline at around 40~d past maximum in $V$ and
$R$ and at $\sim$ 60~d in $I$.  This is evident in our light curves for SN
1993Y, with the change in slope occurring some time between October 1 ($\sim$
36~d after maximum) and October 18 ($\sim$ 53~d after maximum) in $V$ and $R$,
and around October 27 ($\sim$ 62~d after maximum) in $I$.  The data after the
onset of slow decline was fit to a line using a weighted least-squares fit
algorithm (Press \etal~1992).  The corresponding rates of slow decline for
SN~1993Y are shown in Table~7, along with a comparison to the rates of slow
decline for two normal SNe~Ia, SN~1980N (decline rates starting from 42~d past
maximum; Hamuy \etal~1991) and SN~1989B (decline rates from 42~d to 131~d past
maximum; Wells \etal~1994).  As is evident, the results for SN~1993Y are in
agreement with both SNe~1980N and 1989B.

We adopt August 28 as the date SN~1993Z reached maximum brightness; thus, the
observations presented here are from 26~d to 258~d past maximum.  By fitting to
different segments of our data in $R$, we determine that a change in the rate
of decline occurred between September 26 ($\sim$ 29~d after maximum) and
October 22 ($\sim$ 55~d after maximum), which is the case for normal Type Ia
SNe.  Using the same method described above and fitting only to the data after
the onset of the slow linear magnitude decline, the rates for SN~1993Z are
shown in Table~7.  As in the case of SN~1993Y, the rates of slow decline for
SN~1993Z are in agreement with the respective values of the two normal SNe~Ia,
SNe~1980N and 1989B.

The dates of maximum $B$-band brightness of SNe~1993ae, 1994M, 1994Q, and
1994ae were determined from their light curves to be 1993 November 2.1,
1994 May 3.6, 1994 May 28.9, and 1994 November 29, respectively (Riess 1996);
hence, our first observation of SN~1994ae is just 2~d after $B$ maximum.  Those
of SNe~1994B and 1994C were determined from their spectra to be 1994 January 27
and 1994 February 28, respectively (Riess \etal 1999).

Sadakane \etal (1996) present extensive light curves in $VRI$ during the first
100~d for SN~1995D, as well as several spectra near maximum brightness.  They
determine $V$ maximum to be on February 21.5.  It is also seen from their light
curves that $I$ maximum occurs prior to the date of $V$ maximum.  Comparison of
our data with their photometry at about the same epoch shows very good
agreement.  At the first epoch in common, our data in $V$ are fainter by 0.01
mag, which is well within the uncertainties.  Sadakane \etal do not have $R$
and $I$ data at this epoch.  However, they have data 2~d prior and 3~d after
this common epoch, and our $R$ and $I$ data do lie between these points.  At
the second epoch in common, our $V$ and $R$ data are brighter by $\le$ 0.07
mag, and our $I$ data is fainter by 0.21 mag, all within a few times the
uncertainties.  Riess (1996) gives the date of $B$ maximum to be February 21.3.

\subsection{Type IIn Supernova 1994Y}

Early-time spectra of SN~1994Y reveal Balmer emission lines on a wide base,
while at later times the Balmer lines broaden; at all times there is an absence
of narrow absorption lines seen in some other Type~IIn supernovae (Filippenko
1997).  Figure 14 shows the light curves for SN~1994Y.  From the paucity of
data in $B$, it is difficult to say how long SN~1994Y stayed near maximum, but
it is apparent that SN~1994Y had a ``plateau''-like phase near maximum with a
duration of about 30~d in $V$, 40~d in $R$, and 20~d in $I$.  This plateau's
appearance, however, differs from that seen in conventional SNe~II-P (Doggett
\& Branch 1985).

An estimate of the times and magnitudes of peak brightness was made
by eye using the data between JD 2,449,580 and 2,449,700; Table~8
lists the results.  The uncertainty was
estimated from the breadth of the maximum and the abundance of data points near
the peak.  In $B$, the light curve seems to indicate a rapid rise and fall.  In
$V$ and $R$, it reaches the ``plateau'' around (or slightly before) JD
2,449,600 and only gradually rises and falls.  In particular, there is no clear
indication of the time of $V$ maximum, which may have occurred in the gap
between our observations.  $R$ maximum seems to have occurred late in the
plateau.  In $I$, there is a more distinct peak.

It is evident from the $R$ and $I$ light curves that there is a change in the
post-maximum rate of decline around JD 2,449,800 --- the decline rates become
steeper.  There is a similar change in $V$ at about JD 2,449,860.  We fit a
linear decline to the data in $V$ after JD 2,449,700 but prior to 2,449,860,
and in $R$ for all data after JD 2,449,800.  The calculated rates of decline
are shown in Table~7, along with the decline rates for another Type~IIn,
SN~1988Z (rate from 110~d to 450~d past maximum; Turatto et al.~1993), and a
peculiar Type~II, SN~1993J (rate from about 200~d to 400~d after maximum;
Richmond et al.~1996).  SN~1994Y demonstrates a much more rapid decline than
SN~1988Z, but one that is on the same order of magnitude as SN~1993J.

Two indications imply that the mass of the ejecta in SN~1994Y was small.  The
brightness of SN~1994Y remained at a plateau for only several weeks (instead of
several months, as do most SNe II-P), which suggests a small ejecta mass, since
the duration of the plateau phase depends on how long the effective photosphere
takes to work its way back through the hydrogen envelope of the star into the
inner regions.  A short phase indicates that the hydrogen envelope is not
massive (or that the SN was expanding much more quickly than usual).  Also,
Barbon \etal (1995) and Richmond \etal (1996) indicate in the case of SN~1993J
that the small amount of mass ejected, implied from its double peak, leads to a
decrease in the efficiency of $\gamma$-ray trapping; this generates a faster
rate of decline than the 0.98 mag per 100~d resulting from the radioactive
decay of $^{56}$Co, which powers the late-time light curves of normal
Type~II-P SNe.  The similarity in the decline rates of SN~1994Y and SN~1993J,
again, suggests that the ejecta of SN~1994Y were also small in mass.

Finally, the $B$ and $V$ light curves exhibit another change in their rates of
decline beginning around JD 2,449,925, leveling off to about 0.14--0.15 mag per
100~d.  This is similar to the very late-time decline rate of 0.15--0.16 mag
per 100~d seen in SN~1988Z between 730~d and 1150~d (Turatto \etal 1993), which
is a period well after our observations of SN~1994Y.  We suspect that
interaction of the ejecta with circumstellar gas is providing additional light,
as in the case for SN~1988Z.  However, the flattening of the light curves
suggests that this process becomes more pronounced at an earlier time in
SN~1994Y, which again may be a result of the small ejecta mass.

\section{Optical Color Curves}

Figure 15 shows the color curves of the nine Type~Ia SNe.  Note that these
values have not been corrected for reddening.  There are large variations in
the late-time $R - I$ template color curve given by Riess, Press, \& Kirshner
(1996); we have chosen a typical one, but there is a substantial difference
between our observations and this template.  SN~1993Y may be more reddened than
most of the other SNe~Ia presented here.  The proximity of SN~1994C to the host
galaxy's nucleus and the SN's very red color at early times suggest that it
suffered significant reddening or was intrinsically red.  Comparisons of our
colors for SN~1995D with the color curves from Sadakane \etal (1996) show good
agreement.

Figure 16 presents the color curves of SN~1994Y.  SN~1988Z had an initial $B -
V$ $\approx$ 0.4 mag, followed by a leveling off at later times to $B - V$
$\approx$ 0.64 mag (Turatto \etal 1993).  Due to the ambiguity of the time of
maximum for SN~1994Y, it is difficult to indicate the colors at peak
brightness.  However, at the assumed time of $B$ maximum, $B - V$ is between
0.12 and 0.25 mag.  Furthermore, $B - V$ at later times also seems to be
approaching a plateau around 0.5 mag.  $V - R$ appears to increase
monotonically, almost linearly, with time.  $R - I$, on the other hand, is
$-$0.15 mag about two weeks prior to $B$ maximum, reddens to a peak at or above
0.27 mag some time after two months, then turns blueward for the duration of
our observations, ending a year from $B$ maximum at $-$0.78 mag.  The $V - R$
and $R - I$ color curves at late times show the growing strength of $R$
relative to $V$ and $I$, thus suggesting that the H$\alpha$ line is getting
stronger relative to the continuum, and optical spectra confirm this conclusion
(Filippenko 1997; Matheson \& Filippenko 1999, private
communication).\footnote{The H$\alpha$
equivalent width is 1200~\AA\ on 1995 January 25, $\sim 2100$~\AA\ on April 22,
and even larger at later times.} This explains the lack of a
change in the rate of decline at late times of the $R$ light curve, since most
of the light is concentrated in the H$\alpha$ emission.

\section{Absolute Magnitudes}

We can calculate the absolute magnitude of SN~1994Y, for which we have data
near maximum light.  Burstein \& Heiles (1984) determine the Galactic
extinction to NGC~5371 (the host of SN~1994Y) to be $A_{B}$ = 0.00 mag.  Using
the distance modulus to NGC~5371, $\mu = 32.89$ mag (Tully 1988), the resulting
absolute magnitudes are shown in Table~8.  Peak $B$ brightness is $M_{B,max} =
-18.37$ mag.  Average Type~II supernovae have $M_{B,max} = -16.89$ mag, but the
dispersion is quite large, $\sigma = 1.35$ mag (Miller \& Branch 1990).  Like
SN~1988Z, which has a $M_{B,max} \leq -18.0$ mag (Turatto \etal 1993), SN~1994Y
is more luminous than the typical SN~II but not abnormally so.

\section{Conclusions}

We presented $BVRI$ photometry of the Type~IIn SN~1994Y and the Type Ia
SNe~1993Y, 1993Z, 1993ae, 1994B, 1994C, 1994M, 1994Q, 1994ae, and 1995D.  All
observations were done using telescopes of diameter 1~m or less.  Light curves
of SN~1993Y and SN~1993Z exhibit late-time rates of decline very similar to
those of typical SNe~Ia.  The color curves of SN~1993Y suggest that it suffers
more reddening than most of the other SNe~Ia presented here.  SN~1994C may also
have been significantly reddened or was intrinsically red.  The light curves of
SN~1994Y show that it remained at a plateau near maximum light for several
weeks, with no definitive date of peak brightness.  The light curves also
suggest that the ejecta mass was small.  The estimated absolute magnitude of
SN~1994Y shows that it, like the SN~IIn~1988Z, was slightly more luminous than
normal SNe~II.

\acknowledgments

We are grateful to A. J. Barth, B. Leibundgut, and D. Schlegel for their
assistance in obtaining some of the data.  We would also like to express our
appreciation to A. G. Riess for determining the ages of some of the SNe~Ia and
for providing helpful comments. The thoughtful and extensive suggestions of the
anonymous referee led to substantial improvements in this paper.  The research
of A.V.F.'s group was supported by NSF grants AST-8957063, AST-9115174,
AST-9417213, and AST-9987438, as well as by NASA through grants GO-6043 and
GO-6584 from the Space Telescope Science Institute (which is operated by AURA,
Inc., under NASA contract NAS 5-26555). A.V.F. also thanks the Guggenheim
Foundation for a Fellowship.  We are grateful to Sun Microsystems,
Inc. (Academic Equipment Grant Program) and Photometrics, Ltd. for equipment
donations to the telescopes (BAIT) at Leuschner Observatory.  The supernova
search and follow-up efforts with BAIT were the precursors to (and testbeds of)
the Lick Observatory Supernova Search with the Katzman Automatic Imaging
Telescope (Li et al. 2000; Filippenko et al. 2001); continued funding from the
Sylvia and Jim Katzman Foundation is greatly appreciated.

Note added in proof. ---
P. Garnavich, A. Noriega-Crespo, \& A. Moro-Martin (1996, IAU Circ., 6314)
point out that SN 1994Y exhibited a strong infrared excess in data obtained
about 17 months after discovery. This excess could be emission from dust
formed in the ejecta, or it may be an infrared echo from an existing dust
shell.


\clearpage

\setcounter{figure}{0}

\begin{figure}
\caption{SN~1993Y, its host galaxy UGC~2771, and field stars in the $V$ band,
imaged on 1993 November 16. In Figures 1--10 of this paper, the SN or its 
position is marked with an arrow.}
\end{figure}


\begin{figure}
\caption{SN~1993Z, its host galaxy NGC~2775, and field stars in the $I$ band,
imaged on 1993 November 15.}
\end{figure}


\begin{figure}
\caption{SN~1993ae, its host galaxy UGC~1071, and field stars in the $V$ band,
imaged on 1994 January 11.}
\end{figure}


\begin{figure}
\caption{SN~1994B, its host galaxy, and field stars in the $V$ band,
imaged on 1994 February 5. The SN may not be detectable in the printed image.}
\end{figure}


\begin{figure}
\caption{Template image of SN~1994C, its host galaxy, and field stars in
the $I$ band, obtained on 1997 April 12. The SN is no longer detectable in 
this image.}
\end{figure}


\begin{figure}
\caption{SN~1994M, its host galaxy NGC~4493, and field stars in the $V$ band,
imaged on 1994 May 14.}
\end{figure}


\begin{figure}
\caption{SN~1994Q, its host galaxy, and field stars in the $V$ band,
imaged on 1994 July 12.}
\end{figure}


\begin{figure}
\caption{SN~1994ae, its host galaxy NGC~3370, and field stars in the $I$ band,
imaged on 1995 March 30.}
\end{figure}


\begin{figure}
\caption{SN~1995D, its host galaxy NGC~2962, and field stars in the $R$ band,
imaged on 1995 March 28.}
\end{figure}


\begin{figure}
\caption{SN~1994Y, its host galaxy NGC~5371, and field stars in the $R$ band,
imaged on 1995 March 30.}
\end{figure}

\clearpage

\begin{figure}
\plotone{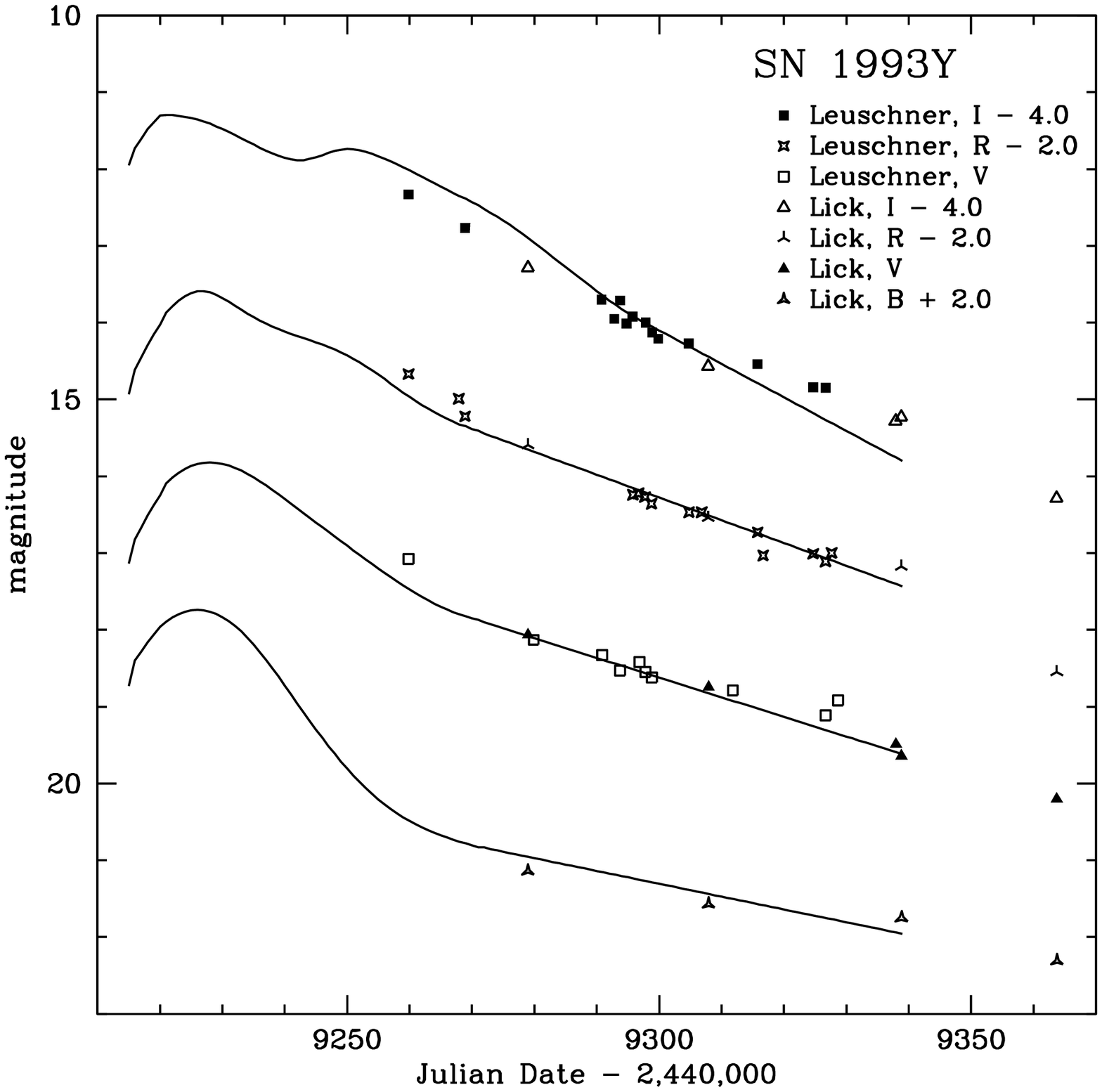}
\caption{$BVRI$ light curves of the Type~Ia SN~1993Y.  Plotted as solid
lines are template Type Ia SN light curves given by Riess \etal (1996).}
\end{figure}

\clearpage

\begin{figure}
\plotone{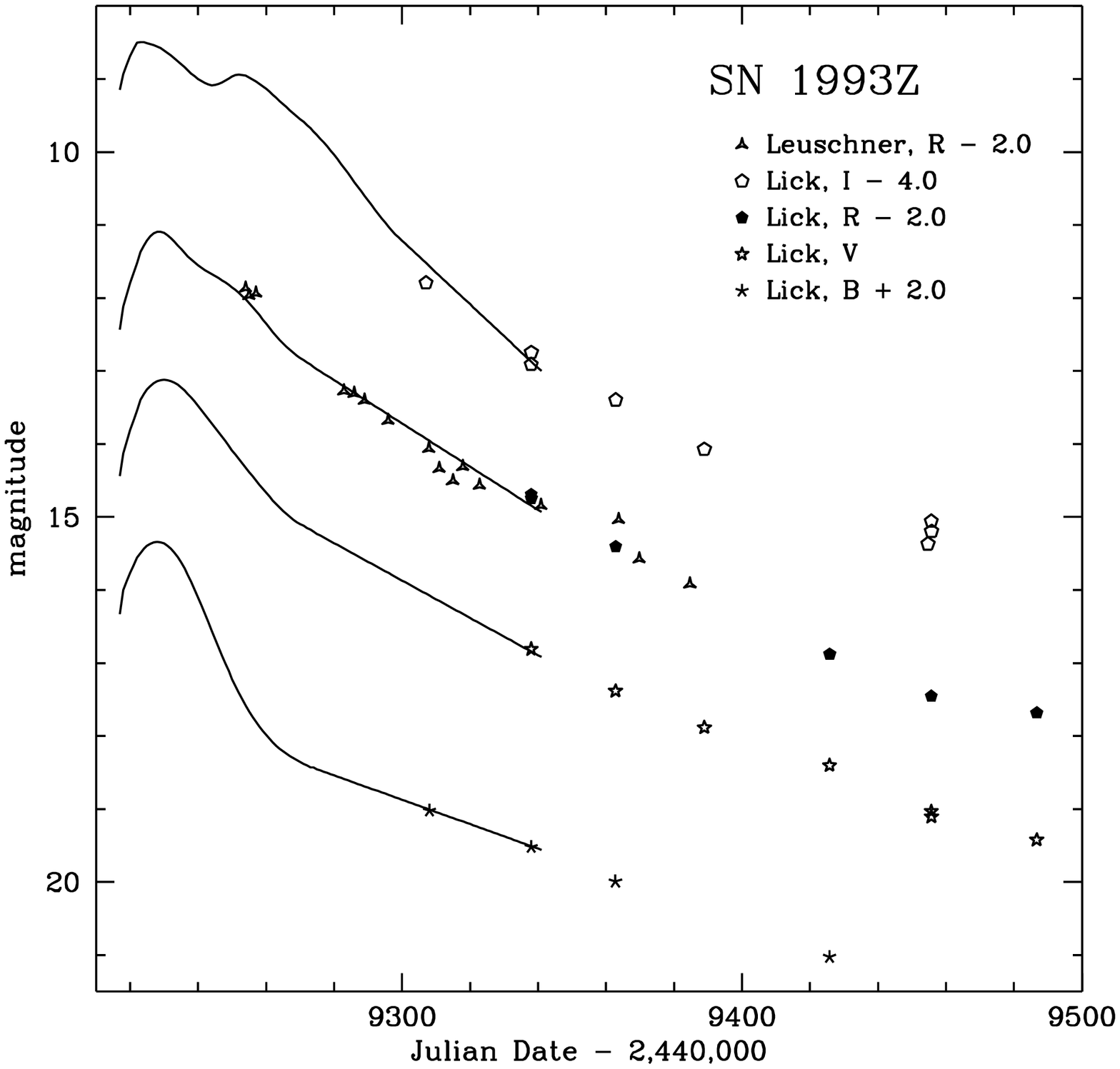}
\caption{$BVRI$ light curves of the Type~Ia SN~1993Z.  Plotted as solid
lines are template Type Ia SN light curves given by Riess \etal (1996).}
\end{figure}

\clearpage

\begin{figure}
\plotone{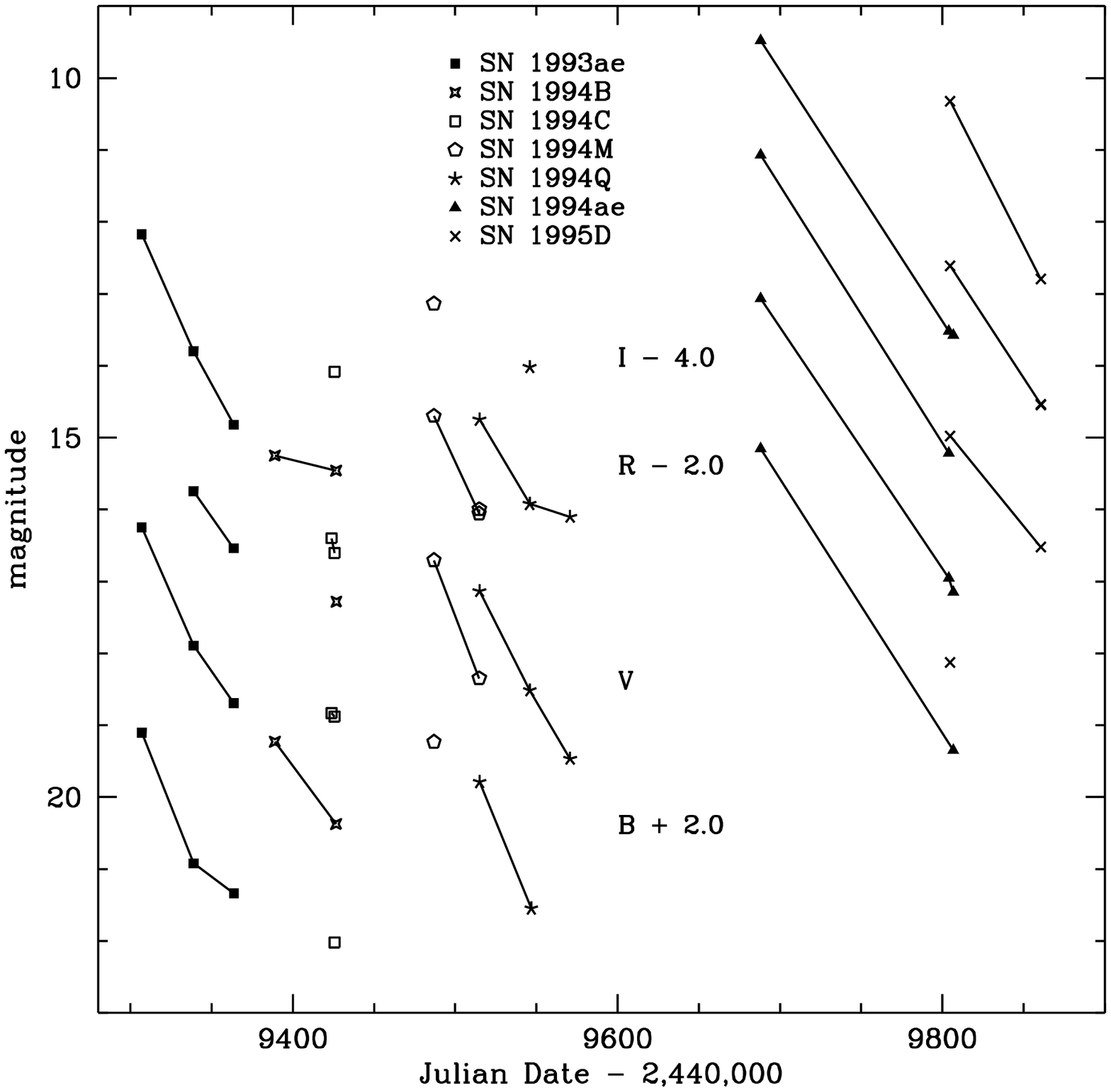}
\caption{$BVRI$ light curves of Type~Ia supernovae.
The line segments connecting the points have no significance other than
to aid in distinguishing between different objects and in indicating
general trends.  There are no $B$ measurements for SN~1994B.}
\end{figure}

\clearpage

\begin{figure}
\plotone{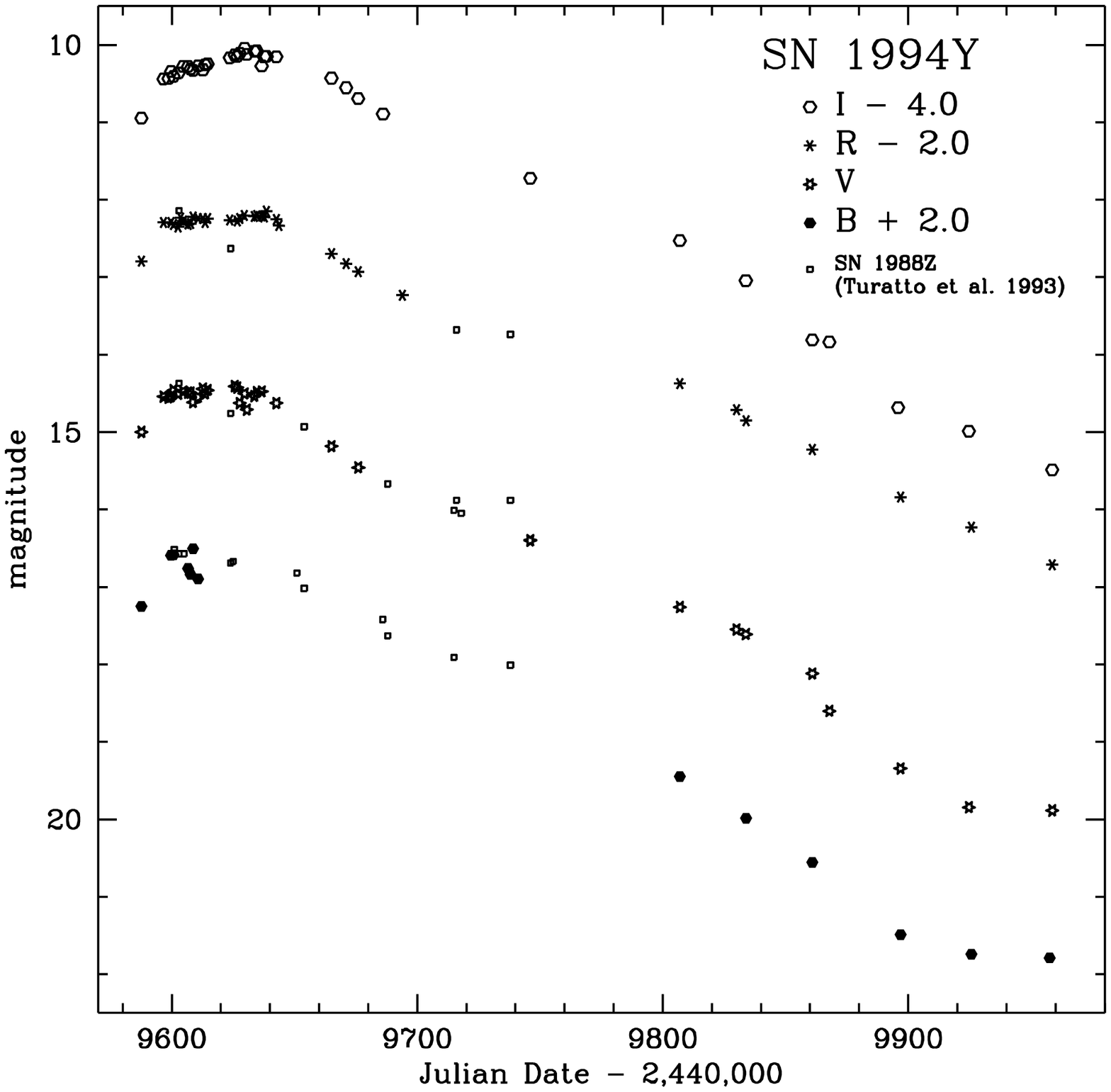}
\caption{$BVRI$ light curves of the Type~IIn SN~1994Y.  Shown for comparison
are the curves for SN 1988Z.}
\end{figure}

\clearpage

\begin{figure}
\plotone{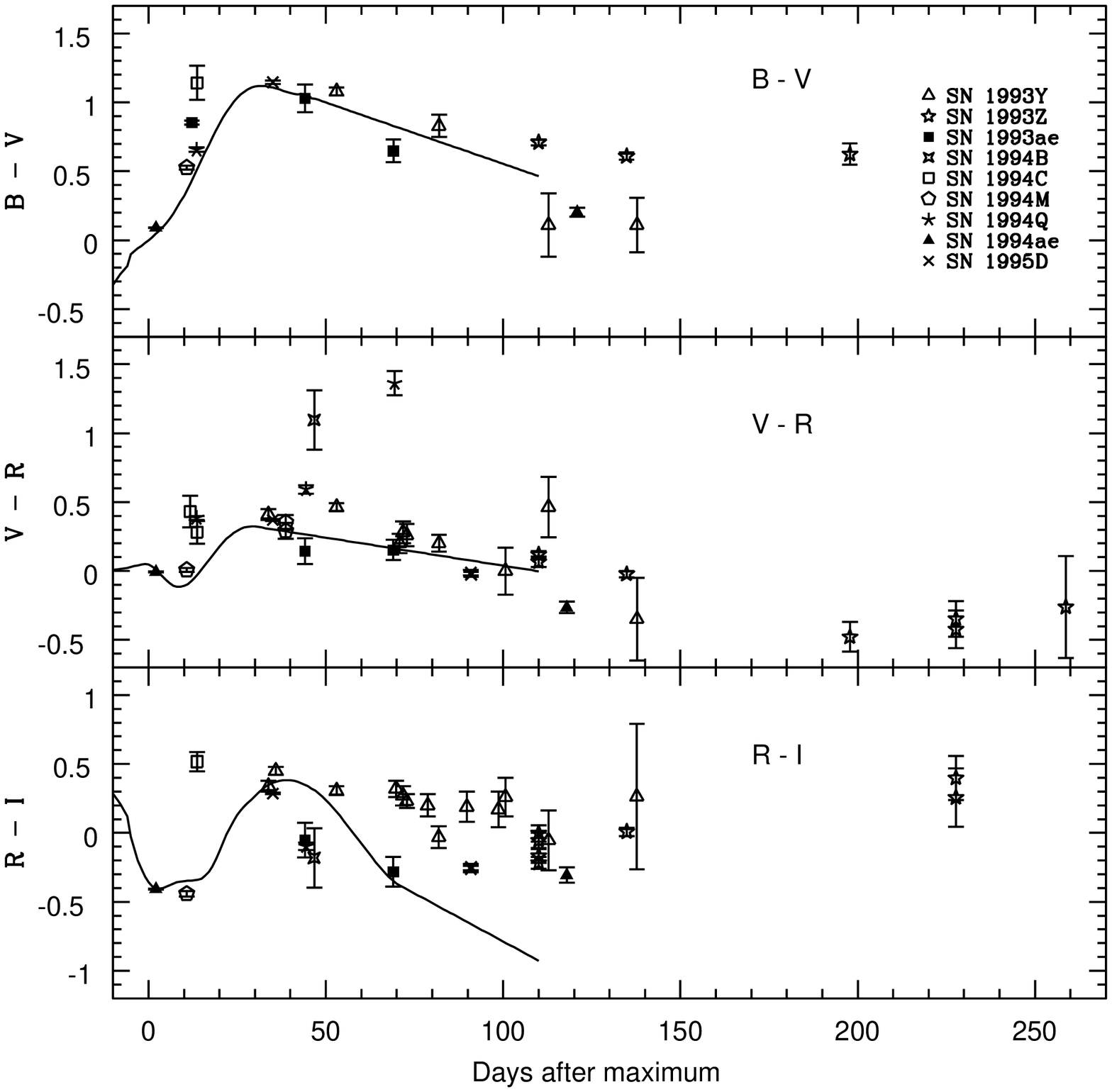}
\caption{$B - V$, $V - R$, and $R - I$ color curves of
Type~Ia~SNe.  Plotted as solid lines are standard SNe~Ia color
curves given by Leibundgut (1988) for $B - V$ and Riess \etal (1996)
for $V - R$ and $R - I$.  There are large variations at late times
in the $R - I$ curves given by Riess \etal (1996), so a typical curve
was chosen.}
\end{figure}

\clearpage

\begin{figure}
\plotone{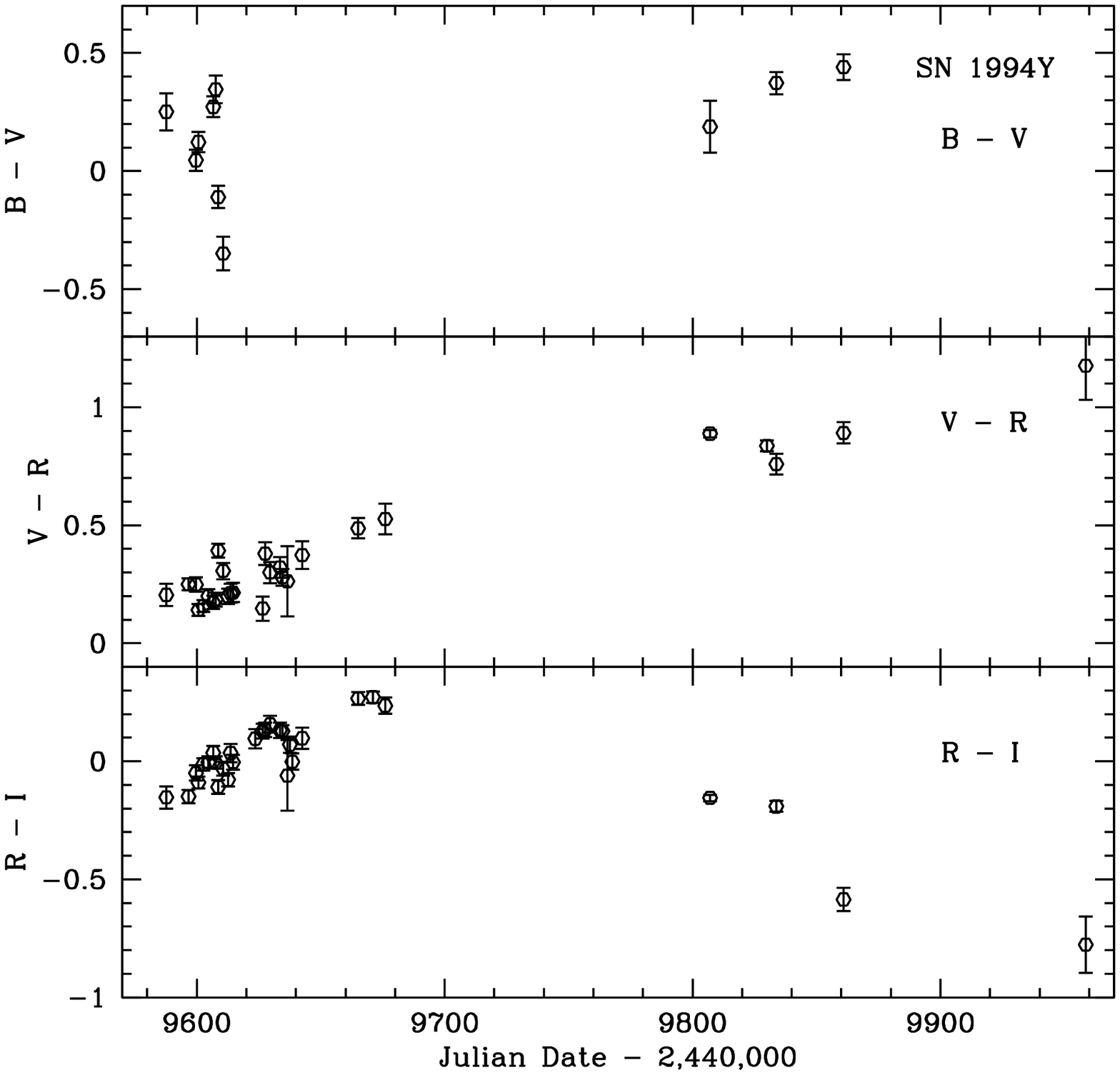}
\caption{$B - V$, $V - R$, and $R - I$ color curves of the Type~IIn SN~1994Y.}
\end{figure}

\clearpage

\begin{deluxetable}{cccccc}
\tablecolumns{6}
\tablecaption{Photometric solutions to transformation equations
 \label{tab:solution}}
\tablehead{\colhead{} & \multicolumn{3}{c}{coefficient}
 & \colhead{photometric} & \colhead{\# of Landolt (1992)} \\
 \colhead{Filter} & \colhead{1} & \colhead{2} & \colhead{3}
 & \colhead{night used} & \colhead{fields observed}}
\startdata
\\ \multicolumn{6}{c}{SN 1993Y, 1993ae, \& 1994Q} \\ \\
 b & 4.49 (02) & 0.31 (01) & $-$0.09 (01) & 97 Sep 05 & 4 \\
 v & 4.10 (02) & 0.20 (01) & {\phmm}0.02 (01) & & \\
 r & 3.95 (01) & 0.14 (01) & {\phmm}0.05 (01) & & \\
 i & 4.03 (03) & 0.10 (01) & $-$0.03 (06) & & \\
\\ \multicolumn{6}{c}{SN 1993Z \& 1994ae} \\ \\
 b & 4.09 (01) & 0.28 (01) & $-$0.06 (00) & 94 Dec 01 & 7 \\
 v & 3.85 (01) & 0.14 (01) & {\phmm}0.02 (01) & & \\
 r & 3.74 (01) & 0.11 (01) & {\phmm}0.04 (01) & & \\
 i & 3.85 (01) & 0.08 (01) & $-$0.02 (01) & & \\
\\ \multicolumn{6}{c}{SN 1994C} \\ \\
 b & 4.55 (04) & 0.25 (03) & $-$0.07 (02) & 97 Apr 12 & 5 \\
 v & 4.04 (03) & 0.25 (02) & {\phmm}0.02 (01) & & \\
 r & 3.92 (03) & 0.18 (02) & {\phmm}0.05 (03) & & \\
 i & 4.08 (03) & 0.09 (02) & $-$0.00 (03) & & \\
\\ \multicolumn{6}{c}{SN 1995D} \\ \\
 b & 4.12 (02) & 0.45 (01) & $-$0.10 (01) & 95 Mar 28 & 7 \\
 v & 4.03 (01) & 0.18 (01) & {\phmm}0.01 (01) & & \\
 r & 3.94 (02) & 0.12 (01) & {\phmm}0.02 (02) & & \\
 i & 4.02 (02) & 0.08 (01) & $-$0.07 (02) & & \\
\\ \multicolumn{6}{c}{SN 1994Y} \\ \\
 b & 4.29 (01) & 0.41 (01) & $-$0.07 (01) & 95 May 23 & 5 \\
 v & 3.98 (02) & 0.28 (01) & {\phmm}0.01 (01) & & \\
 r & 3.86 (01) & 0.24 (01) & {\phmm}0.01 (02) & & \\
 i & 3.98 (04) & 0.17 (01) & $-$0.04 (07) & &
\enddata
\end{deluxetable}

\begin{deluxetable}{ccccc}
\tablecolumns{5}
\tablecaption{Field star apparent magnitudes \label{tab:fieldmag}}
\tablehead{\colhead{star} & \colhead{B} & \colhead{V} & \colhead{R}
 & \colhead{I}}
\startdata
\\ \multicolumn{5}{c}{SN 1993Y} \\ \\
A & 20.17 (29) & 18.88 (10) & 18.90 (19) & 18.38 (24) \\
B & 18.83 (10) & 18.06 (05) & 17.82 (08) & 17.32 (10) \\
C & 18.86 (10) & 18.03 (05) & 17.51 (08) & 16.88 (09) \\
D & 18.77 (10) & 18.02 (05) & 17.57 (08) & 17.10 (09) \\
E & 19.65 (18) & 18.44 (07) & 17.73 (11) & 16.82 (11) \\
F & 17.35 (03) & 16.72 (01) & 16.34 (02) & 15.91 (03) \\
G & 18.65 (08) & 17.50 (03) & 16.74 (04) & 16.04 (05) \\
H & 16.66 (02) & 15.82 (01) & 15.35 (01) & 14.89 (01) \\
I & 16.79 (02) & 16.26 (01) & 15.92 (02) & 15.55 (02) \\
J & 17.28 (02) & 16.33 (01) & 15.76 (02) & 15.20 (02) \\
\\ \multicolumn{5}{c}{SN 1993Z} \\ \\
A & 16.04 (00) & 14.87 (00) & 14.15 (00) & 13.49 (00) \\
B & 16.62 (01) & 15.77 (00) & 15.27 (01) & 14.82 (01) \\
C & 16.65 (01) & 15.79 (00) & 15.31 (01) & 14.86 (01) \\
D & 18.16 (03) & 17.46 (02) & 17.02 (03) & 16.62 (04) \\
E & 20.45 (17) & 18.97 (06) & 17.93 (10) & \nodata    \\
F & 18.94 (05) & 17.83 (02) & 17.11 (03) & 16.57 (04) \\
G & 17.64 (02) & 17.05 (01) & 16.66 (02) & 16.36 (02) \\
\\ \multicolumn{5}{c}{SN 1993ae} \\ \\
A & 18.25 (08) & 17.28 (03) & 16.78 (04) & 16.31 (05) \\
\\ \multicolumn{5}{c}{SN 1994B\tablenotemark{a}} \\ \\
A & 15.60 (02) & 14.96 (02) & 14.55 (02) & 14.22 (02) \\
B & 17.11 (02) & 16.53 (02) & 16.16 (02) & 15.84 (02) \\
C & 16.24 (02) & 15.45 (02) & 14.96 (02) & 14.55 (02) \\
D & 16.12 (02) & 15.58 (02) & 15.23 (02) & 14.94 (02) \\
\\ \multicolumn{5}{c}{SN 1994C} \\ \\
A & 18.82 (10) & 18.06 (05) & 17.69 (08) & 17.19 (10) \\
B & 19.70 (20) & 18.70 (09) & 17.95 (14) & 17.25 (15) \\
C & 20.30 (29) & 18.86 (10) & 17.92 (16) & 17.66 (17) \\
D & 15.78 (01) & 14.91 (00) & 14.40 (00) & 13.95 (01) \\
E & 16.07 (01) & 15.53 (00) & 15.20 (01) & 14.86 (01) \\
F & 13.86 (00) & 13.35 (00) & 13.06 (00) & 12.74 (00) \\
G & 16.30 (01) & 15.38 (00) & 14.88 (01) & 14.40 (01) \\
H & 17.93 (04) & 17.17 (02) & 16.68 (04) & 16.23 (04) \\
I & 17.97 (04) & 16.93 (02) & 16.24 (03) & 15.67 (03) \\
J & 17.71 (04) & 16.99 (02) & 16.60 (03) & 16.17 (04) \\
K & 14.32 (00) & 13.60 (00) & 13.21 (00) & 12.85 (00) \\
L & 14.73 (00) & 14.03 (00) & 13.64 (00) & 13.25 (00) \\
\\ \multicolumn{5}{c}{SN 1994M\tablenotemark{a}} \\ \\
A & 14.20 (02) & 13.19 (02) & 12.68 (02) & 12.32 (02) \\
B & 15.09 (02) & 14.46 (02) & 14.08 (02) & 13.69 (02) \\
\\ \multicolumn{5}{c}{SN 1994Q} \\ \\
A & 15.63 (01) & 15.18 (00) & 14.89 (01) & 14.60 (01) \\
B & 14.77 (00) & 14.22 (00) & 13.87 (00) & 13.54 (00) \\
C & 18.04 (04) & 17.05 (02) & 16.48 (03) & 15.97 (03) \\
\\ \multicolumn{5}{c}{SN 1994ae} \\ \\
A & 18.07 (11) & 17.58 (06) & 17.13 (09) & 16.58 (10) \\
B & 19.06 (26) & 18.57 (14) & 17.31 (20) & 15.84 (21) \\
C & 18.84 (22) & 18.34 (12) & 18.02 (19) & 17.93 (24) \\
D & 18.81 (95) & 20.32 (64) & 18.78 (94) & 18.40 (97) \\
E & 17.18 (05) & 16.53 (02) & 16.11 (04) & 15.66 (04) \\
F & 18.65 (24) & 18.67 (14) & 18.14 (23) & 17.59 (26) \\
G & 19.06 (18) & 17.66 (06) & 16.52 (09) & 14.95 (10) \\
\\ \multicolumn{5}{c}{SN 1995D} \\ \\
A & 18.51 (12) & 17.01 (03) & 16.00 (04) & 14.91 (04) \\
B & 18.22 (10) & 17.21 (04) & 16.61 (05) & 16.15 (06) \\
C & 14.97 (01) & 14.30 (00) & 13.93 (01) & 13.62 (01) \\
D & 15.00 (01) & 14.34 (00) & 13.97 (01) & 13.66 (01) \\
E & 16.64 (03) & 15.87 (01) & 15.44 (02) & 15.08 (02) \\
\\ \multicolumn{5}{c}{SN 1994Y} \\ \\
A & 14.27 (00) & 13.93 (00) & 13.74 (00) & 13.62 (00) \\
B & 14.72 (00) & 14.27 (00) & 13.98 (00) & 13.78 (00) \\
C & 15.81 (01) & 15.35 (00) & 15.08 (01) & \nodata    \\
D & 18.57 (08) & 17.66 (04) & 16.83 (06) & 16.24 (07) \\
E & 18.71 (11) & 18.16 (06) & 17.57 (10) & 17.06 (11) \\
F & 14.77 (00) & 14.19 (00) & 13.82 (00) & 13.53 (00) \\
G & 18.28 (08) & 17.84 (04) & 17.48 (08) & 17.14 (10) \\
H & 16.54 (02) & 15.92 (01) & 15.54 (01) & 15.27 (02)
\enddata
\tablenotetext{a}{Used field star magnitudes from Riess (1996).}
\end{deluxetable}

\begin{deluxetable}{lccccc}
\tablecolumns{6}
\tablecaption{SN~1993Y apparent magnitudes \label{tab:sn93ymag}}
\tablehead{\colhead{UT Date} & \colhead{JD} & \colhead{B} & \colhead{V}
 & \colhead{R} & \colhead{I}}
\startdata
\\ \multicolumn{6}{c}{SN 1993Y (Leuschner)} \\ \\
93 Sep 29.34 & 2449259.84 & \nodata    & 17.08 (04) & 16.67 (02) & 16.33 (03) \\
93 Oct 07.42 & 2449267.92 & \nodata    & \nodata    & 16.99 (08) & \nodata    \\
93 Oct 08.41 & 2449268.91 & \nodata    & \nodata    & 17.22 (02) & 16.77 (02) \\
93 Oct 19.40 & 2449279.90 & \nodata    & 18.13 (05) & \nodata    & \nodata    \\
93 Oct 30.30 & 2449290.80 & \nodata    & 18.33 (10) & \nodata    & 17.70 (10) \\
93 Nov 01.33 & 2449292.83 & \nodata    & \nodata    & \nodata    & 17.95 (08) \\
93 Nov 02.25 & 2449293.75 & \nodata    & 18.53 (12) & \nodata    & 17.71 (09) \\
93 Nov 03.25 & 2449294.75 & \nodata    & \nodata    & \nodata    & 18.01 (08) \\
93 Nov 04.32 & 2449295.82 & \nodata    & \nodata    & 18.24 (03) & 17.92 (05) \\
93 Nov 05.32 & 2449296.82 & \nodata    & 18.42 (06) & 18.22 (04) & \nodata    \\
93 Nov 06.34 & 2449297.84 & \nodata    & 18.55 (06) & 18.27 (05) & 18.00 (05) \\
93 Nov 07.33 & 2449298.83 & \nodata    & 18.62 (07) & 18.36 (04) & 18.13 (03) \\
93 Nov 08.35 & 2449299.85 & \nodata    & \nodata    & \nodata    & 18.21 (05) \\
93 Nov 13.27 & 2449304.77 & \nodata    & \nodata    & 18.47 (04) & 18.27 (07) \\
93 Nov 15.30 & 2449306.80 & \nodata    & \nodata    & 18.47 (04) & \nodata    \\
93 Nov 20.31 & 2449311.81 & \nodata    & 18.79 (09) & \nodata    & \nodata    \\
93 Nov 24.27 & 2449315.77 & \nodata    & \nodata    & 18.73 (07) & 18.54 (09) \\
93 Nov 25.20 & 2449316.70 & \nodata    & \nodata    & 19.03 (09) & \nodata    \\
93 Dec 03.17 & 2449324.67 & \nodata    & \nodata    & 19.01 (08) & 18.84 (10) \\
93 Dec 05.17 & 2449326.67 & \nodata    & 19.11 (15) & 19.11 (08) & 18.85 (11) \\
93 Dec 06.17 & 2449327.67 & \nodata    & \nodata    & 19.00 (07) & \nodata    \\
93 Dec 07.17 & 2449328.67 & \nodata    & 18.92 (17) & \nodata    & \nodata    \\
\\ \multicolumn{6}{c}{SN 1993Y (Lick)} \\ \\
93 Oct 18.50 & 2449279.00 & 19.14 (02) & 18.06 (02) & 17.60 (02) & 17.29 (03) \\
93 Nov 16.40 & 2449307.90 & 19.57 (07) & 18.74 (05) & 18.54 (04) & 18.57 (07) \\
93 Dec 16.43 & 2449337.93 & \nodata    & 19.49 (21) & \nodata    & 19.28 (22) \\
93 Dec 17.34 & 2449338.84 & 19.75 (14) & 19.64 (18) & 19.18 (12) & 19.23 (18) \\
94 Jan 11.24 & 2449363.74 & 20.31 (11) & 20.20 (16) & 20.55 (25) & 20.29 (47)
\enddata
\end{deluxetable}

\begin{deluxetable}{lccccc}
\tablecolumns{6}
\tablecaption{SN~1993Z apparent magnitudes \label{tab:sn93zmag}}
\tablehead{\colhead{UT Date} & \colhead{JD} & \colhead{B} & \colhead{V}
 & \colhead{R} & \colhead{I}}
\startdata
\\ \multicolumn{6}{c}{SN 1993Z (Leuschner)} \\ \\
93 Sep 23.52 & 2449254.02 & \nodata    & \nodata    & 13.87 (07) & \nodata    \\
93 Sep 24.52 & 2449255.02 & \nodata    & \nodata    & 13.96 (08) & \nodata    \\
93 Sep 26.53 & 2449257.03 & \nodata    & \nodata    & 13.94 (08) & \nodata    \\
93 Oct 22.45 & 2449282.95 & \nodata    & \nodata    & 15.27 (17) & \nodata    \\
93 Oct 25.44 & 2449285.94 & \nodata    & \nodata    & 15.31 (17) & \nodata    \\
93 Oct 28.45 & 2449288.95 & \nodata    & \nodata    & 15.40 (18) & \nodata    \\
93 Nov 04.42 & 2449295.92 & \nodata    & \nodata    & 15.68 (20) & \nodata    \\
93 Nov 16.39 & 2449307.89 & \nodata    & \nodata    & 16.06 (24) & \nodata    \\
93 Nov 19.48 & 2449310.98 & \nodata    & \nodata    & 16.34 (26) & \nodata    \\
93 Nov 23.46 & 2449314.96 & \nodata    & \nodata    & 16.50 (28) & \nodata    \\
93 Nov 26.38 & 2449317.88 & \nodata    & \nodata    & 16.31 (26) & \nodata    \\
93 Dec 01.36 & 2449322.86 & \nodata    & \nodata    & 16.57 (28) & \nodata    \\
93 Dec 19.37 & 2449340.87 & \nodata    & \nodata    & 16.84 (31) & \nodata    \\
93 Jan 11.28 & 2449363.78 & \nodata    & \nodata    & 17.03 (33) & \nodata    \\
93 Jan 17.29 & 2449369.79 & \nodata    & \nodata    & 17.57 (38) & \nodata    \\
93 Feb 01.27 & 2449384.77 & \nodata    & \nodata    & 17.92 (42) & \nodata    \\
\\ \multicolumn{6}{c}{SN 1993Z (Lick)} \\ \\
93 Nov 15.57 & 2449307.07 & \nodata    & \nodata    & \nodata    & 15.79 (01) \\
93 Nov 16.57 & 2449308.07 & 17.02 (02) & \nodata    & \nodata    & \nodata    \\
93 Dec 16.52 & 2449338.02 & 17.52 (02) & 16.81 (01) & 16.69 (02) & 16.91 (04) \\
             &            &            &            & 16.74 (03) & 16.75 (05) \\
94 Jan 10.32 & 2449362.82 & 17.99 (02) & 17.38 (02) & 17.40 (02) & 17.40 (02) \\
94 Feb 05.43 & 2449388.93 & \nodata    & 17.88 (03) & \nodata    & 18.07 (05) \\
94 Mar 14.29 & 2449425.79 & 19.02 (06) & 18.40 (05) & 18.88 (10) & \nodata    \\
94 Apr 12.21 & 2449454.71 & \nodata    & \nodata    & \nodata    & 19.37 (42) \\
94 Apr 13.22 & 2449455.72 & \nodata    & 19.03 (09) & 19.45 (10) & 19.06 (12) \\
             &            &            & 19.11 (08) &            & 19.20 (19) \\
94 May 14.22 & 2449486.72 & \nodata    & 19.42 (16) & 19.68 (34) & \nodata
\enddata
\end{deluxetable}

\begin{deluxetable}{lccccc}
\tablecolumns{6}
\tablecaption{SN Ia apparent magnitudes (Lick) \label{tab:snIamag}}
\tablehead{\colhead{UT Date} & \colhead{JD} & \colhead{B} & \colhead{V}
 & \colhead{R} & \colhead{I}}
\startdata
\\ \multicolumn{6}{c}{SN 1993ae} \\ \\
93 Nov 15.34 & 2449306.84 & 17.10 (01) & 16.25 (01) & \nodata    & 16.17 (01) \\
93 Dec 17.21 & 2449338.71 & 18.92 (09) & 17.89 (05) & 17.75 (08) & 17.80 (10) \\
94 Jan 11.14 & 2449363.64 & 19.34 (06) & 18.69 (05) & 18.54 (05) & 18.82 (10) \\
\\ \multicolumn{6}{c}{SN 1994B} \\ \\
94 Feb 05.37 & 2449388.87 & \nodata    & 19.23 (04) & \nodata    & 19.25 (07) \\
94 Mar 15.23 & 2449426.73 & \nodata    & 20.38 (17) & 19.28 (13) & 19.46 (17) \\
\\ \multicolumn{6}{c}{SN 1994C} \\ \\
94 Mar 12.27 & 2449423.77 & \nodata    & 18.83 (07) & 18.40 (09) & \nodata    \\
94 Mar 14.20 & 2449425.70 & 20.02 (11) & 18.88 (06) & 18.60 (06) & 18.08 (04) \\
\\ \multicolumn{6}{c}{SN 1994M} \\ \\
94 May 14.33 & 2449486.83 & 17.23 (01) & 16.71 (01) & 16.70 (01) & 17.14 (02) \\
94 Jun 11.27 & 2449514.77 & \nodata    & 18.35 (04) & 18.06 (04) & \nodata    \\
             &            &            &            & 18.00 (04) &            \\
\\ \multicolumn{6}{c}{SN 1994Q} \\ \\
94 Jun 11.43 & 2449514.93 & 17.79 (01) & 17.13 (01) & 16.75 (01) & \nodata    \\
94 Jul 12.35 & 2449545.85 & \nodata    & 18.51 (02) & 17.92 (02) & 18.02 (02) \\
94 Jul 13.40 & 2449546.90 & 19.55 (07) & \nodata    & \nodata    & \nodata    \\
94 Aug 06.28 & 2449570.78 & \nodata    & 19.46 (07) & 18.10 (05) & \nodata    \\
\\ \multicolumn{6}{c}{SN 1994ae} \\ \\
94 Dec 01.58 & 2449688.08 & 13.15 (00) & 13.06 (00) & 13.07 (00) & 13.48 (00) \\
95 Mar 27.44 & 2449803.94 & \nodata    & 16.95 (02) & 17.22 (03) & 17.52 (04) \\
95 Mar 30.36 & 2449806.86 & 17.35 (02) & 17.15 (03) & \nodata    & 17.58 (03) \\
\\ \\
\\ \multicolumn{6}{c}{SN 1995D} \\ \\
95 Mar 28.33 & 2449804.83 & 16.12 (01) & 14.98 (00) & 14.61 (00) & 14.32 (00) \\
95 May 23.23 & 2449860.73 & \nodata    & 16.52 (01) & 16.53 (02) & 16.80 (02) \\
             &            &            &            & 16.55 (01) &
\enddata
\end{deluxetable}

\begin{deluxetable}{lccccc}
\tablecolumns{6}
\tablecaption{SN~1994Y apparent magnitudes \label{tab:sn94ymag}}
\tablehead{\colhead{UT Date} & \colhead{JD} & \colhead{B} & \colhead{V}
 & \colhead{R} & \colhead{I}}
\startdata
\\ \multicolumn{6}{c}{SN 1994Y (Leuschner)} \\ \\
94 Aug 22.17 & 2449587.67 & 15.25 (07) & 15.00 (04) & 14.79 (03) & 14.95 (04) \\
94 Aug 31.16 & 2449596.66 & \nodata    & 14.54 (02) & 14.29 (02) & 14.44 (02) \\
94 Sep 09.16 & 2449598.66 & \nodata    & 14.55 (02) & \nodata    & 14.44 (03) \\
94 Sep 03.16 & 2449599.66 & 14.59 (04) & 14.54 (02) & 14.30 (02) & 14.34 (03) \\
94 Sep 04.19 & 2449600.69 & 14.58 (04) & 14.46 (02) & 14.32 (01) & 14.41 (02) \\
94 Sep 06.15 & 2449602.65 & \nodata    & 14.51 (02) & 14.35 (02) & 14.36 (02) \\
94 Sep 07.15 & 2449603.65 & \nodata    & \nodata    & 14.23 (02) & \nodata    \\
94 Sep 08.20 & 2449604.70 & \nodata    & 14.48 (02) & 14.28 (01) & 14.28 (02) \\
94 Sep 10.15 & 2449606.65 & 14.76 (04) & 14.49 (02) & 14.32 (02) & 14.28 (02) \\
94 Sep 11.19 & 2449607.69 & 14.84 (05) & 14.49 (03) & 14.31 (01) & 14.31 (02) \\
94 Sep 12.15 & 2449608.65 & 14.50 (04) & 14.62 (02) & 14.22 (02) & 14.33 (02) \\
94 Sep 14.19 & 2449610.69 & 14.90 (06) & 14.55 (03) & 14.24 (02) & 14.27 (02) \\
94 Sep 16.18 & 2449612.68 & \nodata    & 14.44 (03) & 14.24 (02) & 14.32 (02) \\
94 Sep 17.14 & 2449613.64 & \nodata    & 14.50 (03) & 14.29 (02) & 14.26 (03) \\
94 Sep 18.18 & 2449614.68 & \nodata    & 14.46 (04) & 14.24 (02) & 14.25 (02) \\
94 Sep 27.13 & 2449623.63 & \nodata    & \nodata    & 14.26 (03) & 14.17 (03) \\
94 Sep 29.13 & 2449625.63 & \nodata    & 14.41 (03) & \nodata    & 14.13 (02) \\
94 Sep 30.13 & 2449626.63 & \nodata    & 14.42 (05) & 14.28 (02) & 14.15 (02) \\
94 Oct 01.13 & 2449627.63 & \nodata    & 14.62 (04) & 14.24 (02) & 14.11 (02) \\
94 Oct 03.12 & 2449629.62 & \nodata    & 14.50 (04) & 14.20 (02) & 14.05 (03) \\
94 Oct 04.12 & 2449630.62 & \nodata    & 14.71 (07) & \nodata    & 14.12 (04) \\
94 Oct 07.12 & 2449633.62 & \nodata    & 14.53 (04) & 14.21 (02) & 14.08 (02) \\
94 Oct 08.12 & 2449634.62 & \nodata    & 14.49 (03) & 14.21 (02) & 14.08 (02) \\
94 Oct 10.12 & 2449636.62 & \nodata    & 14.48 (13) & 14.21 (07) & 14.27 (14) \\
94 Oct 11.11 & 2449637.61 & \nodata    & \nodata    & 14.22 (02) & 14.15 (03) \\
94 Oct 12.11 & 2449638.61 & \nodata    & \nodata    & 14.15 (02) & 14.15 (03) \\
94 Oct 16.11 & 2449642.61 & \nodata    & 14.63 (05) & 14.25 (03) & 14.15 (03) \\
94 Oct 17.10 & 2449643.60 & \nodata    & \nodata    & 14.33 (03) & \nodata    \\
94 Nov 08.54 & 2449665.04 & \nodata    & 15.18 (04) & 14.70 (02) & 14.43 (02) \\
94 Nov 14.55 & 2449671.05 & \nodata    & \nodata    & 14.82 (02) & 14.55 (02) \\
94 Nov 19.51 & 2449676.01 & \nodata    & 15.46 (06) & 14.93 (02) & 14.69 (02) \\
94 Nov 29.56 & 2449686.06 & \nodata    & \nodata    & \nodata    & 14.89 (02) \\
94 Dec 07.51 & 2449694.01 & \nodata    & \nodata    & 15.23 (03) & \nodata    \\
\\
\\ \multicolumn{6}{c}{SN 1994Y (KPNO)} \\ \\
95 Jan 28.51 & 2449746.01 & \nodata    & 16.40 (03) & \nodata    & 15.72 (02) \\
             &            &            & 16.40 (05) &            &            \\
\\ \multicolumn{6}{c}{SN 1994Y (Lick)} \\ \\
95 Mar 30.53 & 2449807.03 & 17.45 (11) & 17.26 (01) & 16.37 (01) & 16.53 (01) \\
95 Apr 22.51 & 2449830.01 & \nodata    & 17.55 (02) & 16.71 (02) & \nodata    \\
95 Apr 26.37 & 2449833.87 & 17.98 (02) & 17.61 (04) & 16.85 (02) & 17.04 (01) \\
95 May 23.45 & 2449860.95 & 18.56 (07) & 18.12 (03) & 17.23 (04) & 17.81 (04) \\
95 May 30.40 & 2449867.90 & \nodata    & 18.60 (04) & \nodata    & 17.84 (02) \\
95 Jun 27.34 & 2449896.84 & \nodata    & 19.34 (08) & \nodata    & 18.68 (05) \\
95 Jun 28.36 & 2449897.86 & 19.49 (13) & \nodata    & 17.84 (03) & \nodata    \\
95 Jul 25.27 & 2449924.77 & \nodata    & 19.84 (10) & \nodata    & 18.99 (07) \\
95 Jul 26.28 & 2449925.78 & 19.74 (10) & \nodata    & 18.23 (05) & \nodata    \\
95 Aug 27.15 & 2449957.65 & 19.79 (23) & \nodata    & \nodata    & \nodata    \\
95 Aug 28.17 & 2449958.67 & \nodata    & 19.89 (13) & 18.71 (06) & 19.49 (10)
\enddata
\end{deluxetable}

\begin{deluxetable}{lccccc}
\tablecolumns{6}
\tablecaption{Supernova late-time decline rates, mag/100~days
 \label{tab:declinerates}}
\tablehead{\colhead{} & \colhead{$\gamma_{B}$} & \colhead{$\gamma_{V}$}
 & \colhead{$\gamma_{R}$} & \colhead{$\gamma_{I}$} & \colhead{reference}}
\startdata
\\ \multicolumn{6}{c}{Type Ia SNe} \\ \\
SN~1993Y & $1.33 \pm 0.11$ & $2.62 \pm 0.10$ & $2.94 \pm 0.08$
 & $3.89 \pm 0.11$ & \\
SN~1993Z & $1.74 \pm 0.04$ & $1.94 \pm 0.03$ & $2.49 \pm 0.05$
 & $2.72 \pm 0.04$ & \\
SN~1980N & $1.6 \pm 0.3$ & $2.1 \pm 0.3$ & \nodata
 & \nodata & Hamuy \etal 1991 \\
SN~1989B & $1.4 \pm 0.1$ & $2.4 \pm 0.2$ & \nodata
 & \nodata & Wells \etal 1994 \\
\\ \multicolumn{6}{c}{Type II SNe} \\ \\
SN~1994Y & \nodata & $1.44 \pm 0.03$ & $1.59 \pm 0.02$ &
 \nodata & \\
SN~1988Z & 0.47 & 0.44 & 0.18 &
 \nodata & Turatto \etal 1993 \\
SN~1993J & $1.22 \pm 0.08$ & $1.41 \pm 0.07$ & $1.24 \pm 0.04$
 & $1.48 \pm 0.06$ & Richmond \etal 1996
\enddata
\end{deluxetable}

\begin{deluxetable}{ccccc}
\tablecolumns{5}
\tablecaption{SN~1994Y peak magnitudes \label{tab:94ypeak}}
\tablehead{\colhead{} & \colhead{B} & \colhead{V} & \colhead{R} & \colhead{I}} 
\startdata
JD & 2449603.7 $\pm$ 0.5 & 2449617.7 $\pm$ 3.0 &
 2449618.5 $\pm$ 4.0 & 2449629.0 $\pm$ 2.0 \\
UT & Sep 07.20 & Sep 21.20 & Sep 22.00 & Oct 03.50 \\
$m_{max}$ & 14.52 $\pm$ 0.04 & 14.36 $\pm$ 0.05
 & 14.14 $\pm$ 0.02 & 14.04 $\pm$ 0.04 \\
$M_{max}$ & $-$18.37 & $-$18.53 & $-$18.75 & $-$18.85
\enddata
\end{deluxetable}

\end{document}